\documentclass{article}
\usepackage[utf8]{inputenc}
\usepackage[margin=0.6in]{geometry}
\title{\bf { Some positive thoughts about negative absolute temperature}}
\author{Anuradha $\text{Gupta}^{1}$ and Deepak $\text{Jain}^2$ \\
\\
(1) S.G.T.B. Khalsa College\\
University of Delhi\\
Delhi 110007, India\\
\\
(2) Deen Dayal Upadhyaya College\\
(University of Delhi)\\
Sector-3, Dwarka,\\
New Delhi 110078, India\\
\\
email: anuradha@sgtbkhalsa.du.ac.in\\
djain@ddu.du.ac.in}
\date{}
\usepackage{amsmath}
\usepackage{graphicx}
\begin{document}
\large
\maketitle
\begin{abstract}
\large \noindent The concept of negative absolute temperature 
has been widely accepted now and is not just a theoretical curiosity. In this brief report, by combining the formulas used in statistical mechanics and thermodynamics, we
have explained some aspects of negative temperature (both mathematically and graphically) in a two-level system. We believe that these simple calculations may give useful and
concrete insights into negative absolute temperature to undergraduate students.

\end{abstract}

\maketitle

\section{Introduction}
Traditionally the temperature of an ideal gas, consisting of point mass particles, is explained by establishing its relationship with average kinetic energy. For a crystalline solid,  temperature is associated with the vibratory motion of nuclei about their mean positions. These associations have opened up ways to link the thermodynamic properties of a material  to its microscopic atomic structure. Further important insights into the  microscopic world can be obtained by using the   Boltzmann definition of entropy.  This established an association  between the observable quantities of the macroscopic world (thermodynamics) and the microscopic one (statistical mechanics) through the following relationship: $$ S = \, k_B\, ln  \, \Omega $$
Here entropy, $S$, is a macroscopic parameter expressed in terms of a microscopic quantity called multiplicity of a state ($\Omega$) and $k_B$ is the Boltzmann constant. In thermal physics, the concept of temperature is introduced as the relationship between  entropy ($S$) and internal energy ($U$). 

%$$ {1 \over T} = {\left( {\partial S \over \partial U }\right)_{V,N}}$$

$$ \frac{1}{T} = {\left( \frac{\partial S} {\partial U }\right)_{V,N}}$$

where $N$ and $V$ are the number of particles and volume respectively. Hence by using the above definition, temperature  can be rewritten as 
$$ \frac{1}{T} = \, {  k_B} {\left(  \frac {\partial ln \, \Omega}{ \partial U} \right)_{V,N}}
$$

 {\it Temperature remains positive} if entropy is an increasing function of internal energy i.e.  a gas consisting of free particles. In this type of physical system there is a lower bound on energy  and the lower level is more populated than the upper level.This can be understood from the fact that occupation number of each level is proportional  to the Boltzmann factor 
 $$ N_i\, \propto \, e^{-\epsilon_i / k_BT} $$
 So  the normalization factor for these $N_i$, popularly known as the partition function, converges only if there is a lower bound on  $\epsilon_i$  for a positive temperature state.
There exists another class of physical systems where entropy decreases with an increase in the internal energy {\it hence temperature becomes negative}.  In this type of system, there is a finite upper bound to the energy spectrum and the higher energy state is more populated than the lower energy state again to facilitate the convergence of partition function.
This concept of negative absolute  temperature is known since the work done by Onsager on 
vortices  present in the two-dimensional hydrodynamics{\cite{1}}. But experimentally negative temperature  was first observed in a magnetic  nuclear spin system by Purcell and Pound in 1951 {\cite{2}}.  Recently it has been achieved in ultra cold bosons in  optical lattices{\cite{3}}. For the complete history of negative temperature and its implications in  thermodynamics, controversies and recent developments, refer to the following studies {\cite{4,5,6,7,8,9,10,11,12}}.
 \vskip 0.4 cm
 \noindent  Motivated by previous research studies on negative temperature, in this paper,  we elaborate on two important issues related to negative temperature.
First, we evaluate the entropy and internal energy of a finite level system when the  population  of energy levels changes (we consider two levels in this work). The idea is to check what would happen to the temperature of such a system with a gradual change in the number of particles of the two levels. The second objective is to obtain the general expression of $S$ (entropy) in terms of  $U$ (internal energy) and $N$ (number of particles) by using the Lagrange interpolation method. We also believe that the above mentioned points related to negative temperature have not been discussed in adequate detail in the literature. \vskip 0.3 cm
 \noindent The outline of the paper is as follows: In Section 2, we describe the set of equations for calculating the entropy of a two-level system in various configurations of particle distributions. The Lagrange interpolation method along with its application to a finite level system is presented in Section 3. Finally, the results are discussed in Section 4. 
 
\vskip 0.4 cm
\section{ Entropy and internal energy of a finite level system}

In this section we obtain the expressions of entropy and internal energy at various configurations of states defined by the distribution of particles in a  system with two non-degenerate levels. We start with case A; when the lower level of energy $\epsilon_1$ is fully populated with $ N_1 = N$ particles and the upper level of energy $\epsilon_2$ is completely vacant, $N_2 = 0$. Subsequently, we obtain both $S$ and $U$ for different cases when the population of the lower level keeps on decreasing and the occupancy of the higher level goes on increasing. In other words we start with the case when $ N_1 = N$ and $ N_2 = 0$ and cover the entire range of variation of particles until $ N_1 = 0$ and $ N_2 = N$. The variation in entropy with respect to internal energy is plotted in Figure 1 and the corresponding various configurations of particles in  the two levels are plotted in Figure 2. Furthermore, Figure 1 shows that it is symmetric about point $U_D$ which corresponds to the maximum entropy of the given system. In this work we have assumed a two-level system with energy  $ \epsilon_1 \, < \, \epsilon_2$ and $ N_1 + N_2 = N$. 
The details of calculations of $S$ and $U$ for various distribution of particles are as follows:
\begin{figure}[ht!]
\centering
\includegraphics[scale=0.4]{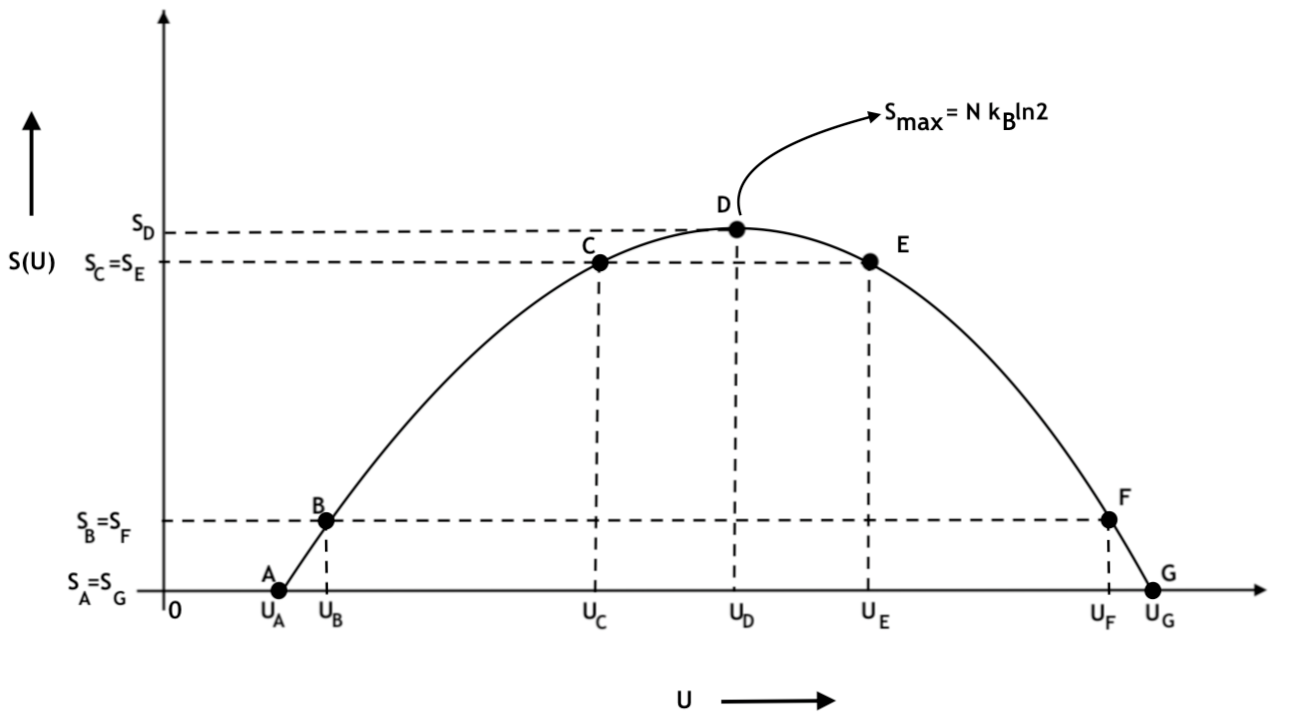}
\caption{ $S$ versus $U$ in a two level system}
\label{6}
\end{figure}

\begin{figure}[ht!]
\centering
\includegraphics[scale=0.4]{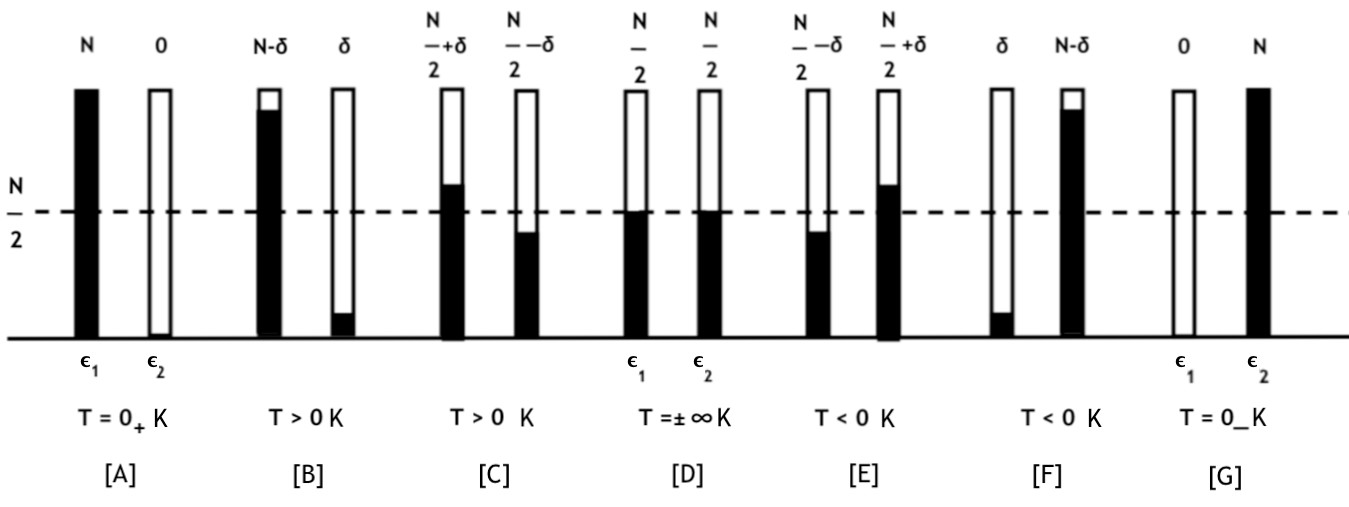}
\caption{Distribution of particles in a two level system where $\epsilon_1 \, < \, \epsilon_2$}
%\label{6}
\end{figure}
\vskip 1 cm
{\bf Case A: Minimum entropy}
\vskip 0.3 cm
$ N_1  = N $ and $N_2 = 0$, Internal energy  is $ U_A = N\epsilon_1$
\vskip  0.4 cm
Entropy at point A can be expressed as:
$$ S_A = k_B \, ln \, \Omega  = \,\,k_B\,\, ln \,{ N! \over {N_1! \, N_2!}} =\,k_B\, ln \,{ N! \over {N! \, 0!}}= k_B\, ln \,(1)\, = \, 0$$  
\vskip 0.5cm

{\bf Case B: Evaluation of $S$ and $U$ at a point close to the minimum entropy configuration}
\vskip 0.3 cm
 $N_1 = {N} - \delta $ and $N_2 = \delta $,
 \vskip 0.3 cm
 Internal energy, $U_B = ({N} - \delta)\epsilon_1 + (\delta)\epsilon_2 = {N} \, \epsilon_1  \, + \,  \delta ( \epsilon_2 - \epsilon_1)$
 \vskip 0.3 cm
 Entropy at point B is equal to: 
 $$ S_B= \, k_B \, ln \,\,{ N! \over { ( {N} - \delta )!\, (\delta )!}} = k_B\,\left[ \, N\, ln \,N - \, N - \,( {N} - \delta )\, ln\, ( {N } -  \delta ) + ( {N } - \delta )\, \right]$$
 
 Since $\delta < < N$,  $ln \, \delta !$ is negligible compared to $ln \,(N - \delta) !$ and $ln \, N !$.

On solving the above equation, we get:

$$ S_B \, = \, k_B\, \left[ \, \delta\, ln \, N - { \delta^2 \over {2 N}}\right]\, = \,  k_B\, \delta\, ln \, N - { k_B\,\delta^2 \over {2 N}}$$

\vskip 0.5 cm
{\bf Case C: Evaluation of $S$ and $U$  at a point close to the maximum entropy configuration}
\vskip 0.3 cm
 $N_1 = {N \over2} + \delta $ and $N_2 = {N \over2} - \delta $.
 \vskip 0.3 cm
 Internal energy, $U_C = ({N \over2} + \delta)\epsilon_1 + ({N \over 2} -\, \delta)\epsilon_2 = {N\over 2} \,( \epsilon_1 + \epsilon_2) \, + \,  \delta ( \epsilon_1 - \epsilon_2)$
 \vskip 0.5 cm
 Entropy at point C is equal to: 
 $$ S_C= \, k_B \, ln \,\,{ N! \over { ( {N \over 2} +  \delta )! ( {N \over 2} -  \delta )!}} = k_B\,\left[ \, N\, ln N - \,( {N \over 2} +  \delta )\, ln\, ( {N \over 2} +  \delta ) - ( {N \over 2} - \delta )\, ln \,( {N \over 2} -  \delta )\right]$$
 
 Using the approximation $ \delta \, << \, { N \over 2}$, we get:
 $$ S_C \, = \, k_B \left[ N \, ln \, N \,  - N \, ln { N \over 2} - {2 {\delta^{2} \over N} } \right ] = \, k_B \, ln\, {2^N}\, - {2\, k_B\, {\delta^{2} \over N} }$$
 
 \vskip 0.5 cm
 
{\bf Case D: Maximum entropy configuration}
\vskip 0.3 cm
$N_1 = { N \over2}$ and $N_2 = { N \over2}$
\vskip 0.3 cm
Internal Energy, $U_D = {N \over2} \, ( \epsilon_1 \, + \epsilon_2)$
\vskip 0.3 cm
Entropy at D is equal to: 
$ S_D =\,  S_{max} = \, k_B \, ln\, {2^N}$

\vskip 0.5 cm

{\bf Case E:  Evaluation of S and U at a point close to the  maximum entropy configuration}
\vskip 0.3 cm
 $N_1 = {N \over2} - \delta $ and $N_2 = {N \over2} + \delta $, 
 \vskip 0.3 cm
 Internal energy, $U_E = ({N \over2} - \delta)\epsilon_1 + ({N \over 2} +\, \delta)\epsilon_2 = {N\over 2} \,( \epsilon_1 + \epsilon_2) \, + \,  \delta ( \epsilon_2 - \epsilon_1)$
 \vskip 0.3 cm
 Entropy at point E is equal to: 
 $$ S_E \,= \, k_B \, ln \,\,{ N! \over { ( {N \over 2} -  \delta )! ( {N \over 2} + \delta )!}} = k_B\,\left[ \, N\, ln N - \,( {N \over 2} +  \delta )\, ln\, ( {N \over 2} +  \delta ) - ( {N \over 2} - \delta )\, ln \,( {N \over 2} -  \delta )\right]$$
 
 Using the approximation $ \delta \, << \, { N \over 2}$, we get:
 $$ S_E \, = \, k_B \left[ N \, ln \, N \,  - N \, ln { N \over 2} - {2 {\delta^{2} \over N} } \right ] = \, k_B \, ln\, {2^N}\, - {2\, k_B\, {\delta^{2} \over N} } = \, S_C$$
 
\vskip 0.5 cm

{\bf Case F: Evaluation of $S$ and $U$ at a point close to the minimum entropy configuration}
\vskip 0.3 cm
 $N_1 = \delta $ and $N_2 =  N - \,\delta $, 
 \vskip 0.3 cm
 Internal energy , $U_F = {N}\epsilon_2  -\,  \delta ( \epsilon_2 - \epsilon_1)$
 \vskip 0.3 cm
 Entropy at point F is equal to: 
 $$ S_F= \, k_B \, ln \,\,{ N! \over {(\delta )! \,( {N} - \delta )! }} = k_B\,\left[ \, N\, ln \,N - \, N - \,( {N} - \delta )\, ln\, ( {N } -  \delta ) + ( {N } - \delta )\, \right] $$

On solving the above equation, we get :

$$ S_F \, = \, k_B\, \left[ \, \delta\, ln \, N - { \delta^2 \over {2 N}}\right] \, = \, S_B$$
\vskip 0.5 cm

{\bf Case G:  Minimum entropy}
\vskip 0.3 cm

$ N_1  = 0 $ and $N_2 = N$, Internal energy  is $ U_G = N\epsilon_2$
\vskip  0.4 cm
Entropy at point G can be expressed as:
$$ S_G = k_B \, ln \, \Omega  = \,\,k_B\,\,ln\, { N! \over {N_1! \, N_2!}} = k_B\, ln (1)\, = \, 0  =\, S_A$$  
\vskip 0.5cm
\noindent The results obtained above are presented in Table {\ref{1T}}. One can clearly observe that as we move from point $A$ to $D$, entropy increases from zero to its maximum value at point $D$ and after that it decreases continuously as we move from $D$ to $G$. On the other hand, internal energy $U$, keeps on increasing continuously from point $A$ to $G$. 
\vskip 0.3 cm
\noindent Another interesting feature of this calculation  is the change in the magnitude of entropy of the system with a variation in the number of particles occupying the two levels of the system. In the vicinity of the minimum entropy position (point A or G), the change in entropy is maximum when we move from $A$ to $B$ (assume $\delta = 1$) or from $ F$ to $ G$.
Similarly the change in entropy is minimum in the vicinity of the maximum entropy position. (e.g. when move from $C$ to $D$ or from $D$ to $E$ and assume $\delta = 1$). Hence the variation in entropy is steep  with a change in the number of particles close to the minimum position. Furthermore the variation in $S$ slows down and becomes almost flat at the maximum position. This behaviour of entropy is symmetrical and hence, one can easily correlate this variation with an inverted parabola graph.
 
\begin{table}[ht]
\centering
\renewcommand{\arraystretch}{3}
\begin{tabular}[b]{| c | c| c| c| c|}\hline
 Case & $N_1$ & $N_2 $& Internal Energy ($U$) & Entropy ($S$) \\ \hline \hline
A & $N$ & 0 &$ N\,\epsilon_1 $& 0\\ \hline
B & $N - \, \delta$& $\delta$ & $ {N} \, \epsilon_1  \, + \,  \delta ( \epsilon_2 - \epsilon_1)$& $ k_B \, \delta\, ln \, N - \, k_B{\delta^2 \over {2 N}}$\\ \hline
C & ${N/2} + \, \delta$&${N/2} - \, \delta$& ${N\over 2} \,( \epsilon_1 + \epsilon_2) \, - \,  \delta ( \epsilon_2 - \epsilon_1)$ & $ k_B \, ln\, {2^N}\, - {2\, k_B\, {\delta^{2} \over N}} $\\ \hline
D & $N/2$ & $N/2$ & ${N\over 2} \,( \epsilon_1 + \epsilon_2)$ & $ k_B \,ln\, 2^N$\\ \hline
E & ${ N / 2 } - \, \delta$ & ${N/2} + \, \delta$ & ${N\over 2} \,( \epsilon_1 + \epsilon_2) \, + \,  \delta ( \epsilon_2 - \epsilon_1)$ & $ k_B \, ln\, {2^N}\, - {2\, k_B\, {\delta^{2} \over N}}  $\\ \hline
F & $\delta$ & $N -\, \delta$ & $ {N} \, \epsilon_2  \, - \,  \delta ( \epsilon_2 - \epsilon_1)$&  $ k_B \, \delta\, ln \, N - k_B\,{ \delta^2 \over {2 N}}$\\ \hline
G & 0 & $N$ & $ N \, \epsilon_2$ & 0\\ \hline
\end{tabular}
\caption{ Brief summary of the parameters obtained in the two-level system}
\label{1T}
`\end{table}

\vskip 0.5cm
\section{ General expression of \,\,$S(U, N)$}

In the previous section, we obtained the values of entropy for a two-level system for various distributions of particles in it. In this section we will focus on the construction of a smooth polynomial\,  (Here $S(U)$)  using the data obtained in the form of $( S_i, U_i)$ in the previous section.
We use the Lagrange interpolation method  to determine the general form of entropy as a function of internal energy. In this method the nth-order polynomial is constructed using $n+1$ data points. 
This can be further written as 
$$ S_{ip}(U) = \, \sum_{i = 1}^{ n} L(U_i)S_i  = L( U_1)\,S_1 + L(U_2)\, S_2 + \, L(U_3) \, S_3 + ..... +\, L(U_n) \, S_n $$ 
\vskip 0.3 cm
\noindent Here, $S_{ip}$ is an interpolating polynomial of degree $n$.  Points $U_1, U_2, ....U_n$ are interpolation points and $L(U)$ is known as the Lagrange polynomial. As mentioned in the previous section, the graph between $S$ and $U$ is an inverted parabola so we assume $S$ to be a second-order polynomial and hence use  three data points. This parabolic variation between $S$ and $U$ is also supported by Masthay and Fannin {\cite{13}}. The three data points used to obtain the general form of $S$ are $(U_A, S_A)$,$(U_C, S_C)$ and  $(U_G, S_G)$ at points $A$, $C$ and $G$ respectively {\footnote{One can also obtain the general form of $S(U)$ by assuming any three points as shown in Figure 1.}}.
For more details of this method see {\cite{14}}.

\vskip 0.4 cm
\noindent Now the general form of $ S_{ip}(U,N)$ in this case becomes:
\begin{equation}
S_{ip}(U, N)  = L(U_A) S_A + L(U_C) S_C + L(U_G) S_G
\label{a}
\end{equation}
\vskip 0.3 cm
as both $ S_A = S_G = 0$, hence $ S_{ip}(U)  = L(U_C) S_C $
\vskip 0.4cm
By definition, the basis polynomial, $L(U_C)$ is written as:

$$ L(U_C) =\, {{(U - U_A) (U - \, U_G)}\over{(U_C - \, U_A) ( U_C - \, U_G)}}$$

After substituting the values of $U_A,\, U_C, \, U_G$ and $S_C$  in Eq.{\ref{a}}, we finally get:

$$S_{ip}(U, N) \,= S_C \, \left[ {( U - N\epsilon_1) \, ( U - N\epsilon_2) }\over { ( { N \over 2} (\epsilon_1 + \epsilon_2) + \delta (\epsilon_1 - \epsilon_2) - \, N \epsilon_1) \left( { N \over 2} (\epsilon_1 + \epsilon_2) + \delta (\epsilon_1 - \epsilon_2) - \, N \epsilon_2 \right)} \right]$$
\vskip 0.5 cm

\begin{equation}
 \boxed{S_{ip}(U, N)\,=  \left[  \,k_B \, ln \,2^N\, - { 2\,k_B\, \delta^2 \over N} \right]\left[ U^2 \, - { N (\epsilon_1 + \epsilon_2) U \, + N^2 \epsilon_1 \epsilon_2 }\over { ( \delta^2 - { N^2 \over 4} ) (\epsilon_1 -\epsilon_2)^2 } \right]}
 \label{b}
\end{equation}

\vskip 0.5 cm

\noindent It is interesting to compare the expression of entropy obtained using the Lagrange interpolation method (Eq.{\ref{b}}) with the exact expression of entropy written in terms of internal energy and the number of particles.  The exact expression of entropy for a two-level system of energies $\epsilon_1$ and $\epsilon_2$ can be written as: 
\begin {equation}
{S_{exact}\over k_B} =  N \,ln \,N \, - {N \over {\epsilon_2 - \epsilon_1}}\left( \epsilon_2 \,\, ln \,\, { {N\epsilon_2 -U }\over{ \epsilon_2 - \epsilon_1}} - \epsilon_1 \,\, ln\, \,{ {U - N\epsilon_1  }\over{ \epsilon_2 - \epsilon_1}}\,\right) 
 + {U \over {\epsilon_2 - \epsilon_1}}\left( \, ln \,{ {N\epsilon_2 -U }\over{ \epsilon_2 - \epsilon_1}} - \,\, ln\,\, { {U - N\epsilon_1  }\over{ \epsilon_2 - \epsilon_1}}\,\right) 
\end{equation}
In a special case, by assuming $ \epsilon_1 = 0 $ and $ \epsilon_2 = \epsilon $, the internal energy of the system is $ U = q \, \epsilon$. The parameter $q$ defines the macrostate $(N-q \, , \, q)$ of the system, which varies as $ 0\le \, q \le N$. The above expression of entropy reduces to
\begin {equation}
{S_{exact} \over k_B} = \left[\,\, N \,ln \,N \, - N \left( \, ln\, {(N -q )} \,\right) 
 + q\left( \, ln \, {(N -q) } - \,\, ln \,{(\, q)}\,\right) \,\,\right]
\end{equation}

\noindent In Fig.3, we plot entropy obtained from the interpolation method and the exact method simultaneously for $N = 1000$ and $ \delta = 1$ with respect to the macrostate parameter $q$. It can be seen that if we keep $\delta$ very small (i.e. $ \delta = 1 $, remaining close to the maximum and minimum entropy positions), both curves match exactly at these points. As expected the difference between the two expressions increases as we move away from the maximum and minimum entropy points. This may be because the exact expression of entropy is a log function of internal energy and the interpolation expression of entropy is a quadratic function of $U$.

\begin{figure}[ht!]
\centering
\includegraphics[scale=0.4]{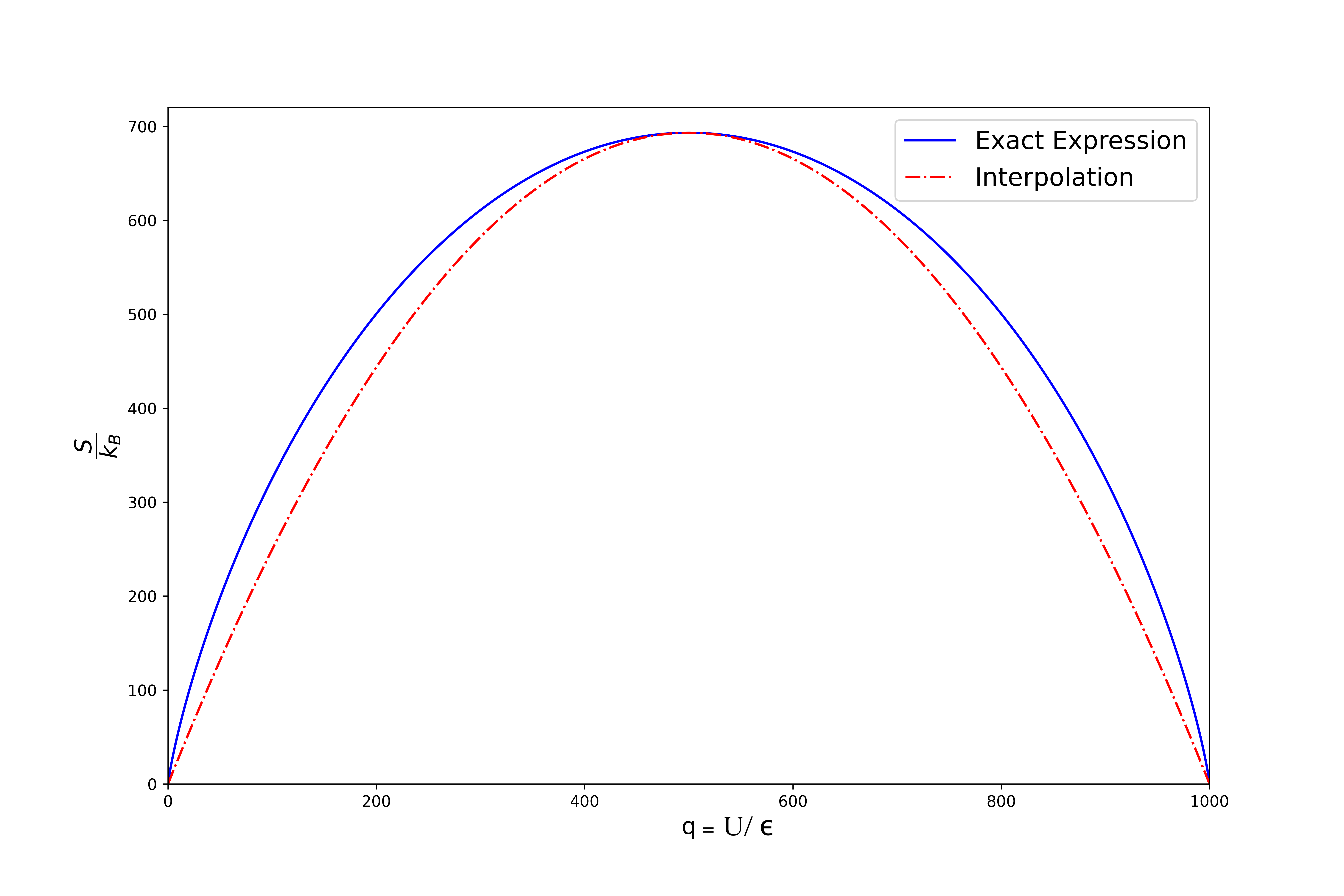}
\caption{ Variation in entropy with respect to various macrostates in a two-level system }
\label{6}
\end{figure}
\section{Discussion}
We investigated two issues related to negative absolute temperature which have not been highlighted explicitly in the literature. Hence
the importance of this study is twofold: 
\vskip 0.5 cm
\noindent {\bf First}, we show  the behaviour of entropy close to the maximum point when the population in both the levels is equal. Similarly the values of entropy are obtained close to the minimum point when one of the levels is fully occupied while the other is vacant. This clearly indicates that the graph between entropy and internal energy is symmetrical. Internal energy increases continuously from point $A$ to $G$ while entropy increases only from point $A$ to $D$ and hence, we obtain positive temperature in this region. While we move from $D$ to $G$, entropy starts decreasing (see Fig.1), but internal energy is still increasing; hence, temperature becomes negative in the right half of the graph. The concept of positive and negative absolute temperature can be understood from Figure 1 by invoking  the Boltzmann distribution. In the left half of the curve ADG, the ratio of the  number of particles in  the two levels given by: $ N_2 /N_1 = \,exp [( \epsilon_1 \, -  \epsilon_2 )/ k_B T ]\, < 1$ and hence temperature is positive. Similarly in the right half of the curve ADG, $N_2/N_1 >1$ and therefore temperature becomes negative.

\vskip 0.5 cm
\noindent{\bf Second}, we establish the  general mathematical relationship between entropy and internal energy for the  two-level system using the Lagrange interpolation method.

\vskip 0.7 cm
\noindent {$\bullet$ It is important to note that  one can obtain  the {\it maximum magnitude  of absolute temperature }  in this finite level system by analysing the  transition from $[{N \over 2 } +1, { N \over 2}  -1]$ to $[ { N\over 2}  , {N\over2} ]$ or from  $[ { N\over 2} , {N\over2} ]$ to $[{N \over 2 } -1, {N \over 2} + 1]$.} In other words, the system attains the maximum value of temperature when it reaches the state of  maximum  entropy [$ {N\over 2}, {N\over 2} $].
\vskip 0.3 cm
\noindent For example, in case C, by assuming $ \delta = 1$ we are close to the maximum entropy state.  So during the transition from 
$[{N \over 2 } + 1, {N \over 2} -1]$ to $[ { N\over 2}  , {N\over2} ]$  or from case C to case D, the change in entropy is 
$$S_D - S_C = { 2\,k_B \over N}$$

and the change in internal energy is given as 
$$U_D - U_C = \epsilon_2 - \epsilon_1$$

and hence the maximum  magnitude of the temperature in this  system is 

$$ T = \, {(\epsilon_2 - \epsilon_1)  \over {2k_B}/N}$$

\noindent In the limit, when $ N \rightarrow \infty$, the maximum magnitude of  temperature approaches $T  =\,+ \,\infty$ K.
\vskip 0.3 cm
\noindent Similarly during the transition from  $[ { N\over 2} , {N\over2} ]$ to $[{N \over 2 } -1, {N \over 2} + 1]$ or from case D to case E, one can easily obtain  $ T \, = \, - \infty$ K in the limit when $  N \rightarrow \infty$.
\vskip 0.7 cm

\noindent $\bullet$ Another interesting feature of such system is that the {\it  minimum  magnitude of absolute temperature} exist at the minimum entropy configuration. This can be easily understood when the system undergoes transition from state [$N$, 0] to [$N-1$, 1] or from [ 1, $N-1$] to [0, $N$]. Hence, 
in case B, if we assume $\delta =1 $, we are close to the minimum entropy position (case A). If the system moves to [$N-1$, 1] state from [$N$,0] state or  transition from case  A to B, the change in entropy is given as:

$$S_B - S_A =\, k_B\, ln{N}\,-{ k_B \over 2N}$$

and the change in internal energy is given as 
$$U_B - U_A = \epsilon_2 - \epsilon_1$$

and hence, the minimum temperature of such a system is 

$$ T = \, {{(\epsilon_2 - \epsilon_1)}  \over {k_B\, ln{N}\,-\, \,{ k_B \over 2N}}} $$

\noindent In the limit, when $ N \rightarrow \infty$, the minimum  magnitude of temperature approaches $T  =\,0_+$ K.
\vskip 0.3 cm
\noindent Similarly during the transition from  $[1 , N- 1 ]$ to $[0, N]$ or from case F to case G, one can easily obtain  $ T \, = \, 0_-$ K in the limit when $  N \rightarrow \infty$. This is incidentally the maximum temperature one can obtain in a two-level system (these points are also discussed at length by assuming various number of particles, see ref.{\cite{13}}).

\vskip 0.3 cm
\noindent To conclude, in this article we explored the behaviour of entropy close to the points where entropy is either maximum or minimum in a finite level system. Furthermore we also obtained the general expression of entropy as a function of internal energy of the given two level system. We hope that this simple calculation will provide highly useful insights and understanding of the concept of negative absolute temperature.


\begin{thebibliography}{99}

\bibitem{1}L. Onsager, "Statistical Hydrodynamics", Suppl. Nuovo Cimento {\bf 6}, 279 (1949).
\bibitem {2}E. M. Purcell and  R.V. Pound, “A nuclear spin system at negative temperature,”
Phys. Rev. {\bf 81}, 279 (1951).
\bibitem{3} S. Braun et al., "Negative Absolute Temperature for Motional Degrees of Freedom", Science, {\bf 339}, 52 ( 2013).
\bibitem{4} M. Baldovin et al., "Statistical mechanics of systems with negative temperature", Phys. Rep., {\bf 923}, 1 (2021).
\bibitem{5} N. F. Ramsey, “Thermodynamics and statistical mechanics at negative
absolute temperatures”,  Phys. Rev. {\bf 103}, 20 (1956).

\bibitem{6} W. G. Proctor, " Negative Absolute Temperature", Sci. Am., {\bf 239}, 90 (1978).
\bibitem{7} I. M. Sokolov, "Not hotter than hot", Nat. Phys. {\bf 10}, 7 (2014).
\bibitem{8} L. D. Carr, "Negative Temperatures ?", Science, {\bf 339}, 42 (2013).
\bibitem{9} J. Dunkel and  S. Hilbert, "Consistent thermostatistics forbids negative absolute temperatures",  Nat. Phys. {\bf 10}, 67 (2014).

\bibitem{10} D. Frenkel and  P. B. Warren, "Gibbs, Boltzmann, and negative temperatures", Am. J. Phys, {\bf 83}, 163 (2015).
\bibitem{11} J. Wisnaik, "Negative Absolute Temperatures, a Novelty" , J. Chem. Educ., {\bf 77}, 518 (2000)
\bibitem{12} R. J. Tykodi, "Negative Kelvin temperatures: Some anomalies and a speculation,”
Am. J. Phys. {\bf 43}, 271 (1975).
\bibitem{13} M.B. Masthay  and H.B. Fannin, " Positive and Negative Temperatures in a Two-Level System:
Thermodynamic and Statistical-Mechanical Perspectives ", J. Chem. Educ., {\bf 82}, 867 (2015).
\bibitem{14} R. L. Burden and J. Douglas Faires, {\it Numerical Analysis}, 9th Edition,( Brooks/ Cole, Cengage Learning 2010)
\end{thebibliography}
\end{document}